\documentclass[prd,aps,showpacs,floats]{revtex4}
\usepackage{amsmath}
\usepackage{amssymb}

\begin{document}

\title{A Bound Quantum Particle in a Riemann-Cartan space with  Topological Defects and Planar Potential}
\author{S.A. Ali${}^{\flat}$, C. Cafaro${}^{\flat}$, S. Capozziello${}^{\sharp}$, Ch. Corda$%
{}^{\dag}$}
\address{${}^{\flat}$Department of Physics, University at
Albany-SUNY, 1400 Washington Avenue, Albany, NY, 12222, USA
\\ ${}^{\sharp}$ Dipartimento di Scienze Fisiche, Universit\`a di Napoli "Federico II" and INFN Sez. di Napoli,
Compl. Univ. Monte S. Angelo, Ed.N, Via Cinthia, I-80126 Napoli,
Italy\\
${}^{\dag}$ INFN Sez. di Pisa and Universit\`a di Pisa, Via F.
Buonarroti 2, I-56127 Pisa,  Italy.}

\begin{abstract}
Starting from a continuum theory of defects, that is the analogous
to three-dimensional Einstein-Cartan-Sciama-Kibble gravity, we
consider a charged particle with spin $\frac{1}{2}$ propagating in
a uniform magnetic field coincident with a wedge dispiration of
finite extent. We assume the particle is bound in the vicinity of
the dispiration by long range attractive (harmonic) and short
range (inverse square) repulsive potentials. Moreover, we consider
the effects of spin-torsion and spin-magnetic field interactions.
Exact expressions for the energy eigenfunctions and eigenvalues
are determined. The limit, in which the defect region becomes
singular, is considered and comparison with the electromagnetic
Aharonov-Bohm effect is made.
\end{abstract}


\maketitle

\vspace{3. mm}


\vspace{5. mm}

\section{Introduction}

The investigation of quantum mechanical systems with non-trivial
boundary conditions is an active field of research. A fertile
ground for such systems is provided by particles moving in a
background space with nonvanishing (positive or negative)
curvature and/or non-trivial topology. By non-trivial topology we
mean multiply connected spaces. Indeed, the prototype representing
the case of non-trivial topology is the well known electromagnetic
Aharonov-Bohm (AB) effect \cite{AB}. In the context of the
electromagnetic AB effect, the effect of topology manifests as the
phase factor in the wavefunction of an electron moving around a
magnetic flux line. The gravitational analogue of this effect has
been investigated in \cite{2, 3, 4}. We are going to investigate a
space with topological defects characterized by vanishing
Riemann-Christoffel curvature and torsion \cite{MTW} everywhere
except on the defects. Such defects arise in gauge theories with
spontaneous symmetry breaking and may have played some significant
role in the formation of large scale cosmological structure. Some
examples in cosmology are domain walls \cite{5}, cosmic strings
\cite{5, 6} and monopoles \cite{7}. Analogues of such defects in
condensed matter physics include vortices in superfluids and
superconductors \cite{8}, domain walls in magnetic materials,
dislocations in solids and disclinations in liquid crystals or
two-dimensional graphite \cite{9}.

Quantum effects on particles moving in a crystalline media with
topological defects have been a subject of investigation since the
early 1950's \cite{10}. In the geometric approach of Katanaev and
Volovich \cite{11}, the theory of defects in solids is translated
into the language of three-dimensional
Einstein-Cartan-Sciama-Kibble gravity \cite{Kibble,Hehl} but it
could be extended  considering more general forms of torsion
\cite{CLS}. This formalism corresponds to viewing the continuous
limit of a crystalline solid in which the defect configuration is
characterized by a non-trivial metric describing a static
three-space. In this way, elastic deformations introduced in the
medium by defects are incorporated into the metric manifold. The
boundary conditions required by the presence of such defects are
accounted for by the associated non-Euclidean metric. In the
continuum limit, that is, for distances much larger than the
Bravais lattice spacing, the theory describes a non-Riemannian
manifold  where curvature and torsion are associated with
disclinations and dislocations in the medium, respectively.

In the case of solid crystals, a topological defect can be thought as
consisting of a core region characterized by the absence of regular lattice
order and an ordered (undistorted) far-field region \cite{16}. Although in
the continuum approximation in which we work the core region is usually
shrunk to a singularity, we choose to relax this condition by smearing the
singularity over a finite region. By virtue of having a finite defect
region, we consider the effects of spin-torsion and spin-magnetic field
interactions on the system.

In this work, we study a charged spin-$\frac{1}{2}$ particle
moving in the space of a defect comprised of a combined screw
dislocation and a wedge disclination. Such a combined defect is
called a wedge dispiration, a defect possessing nonvanishing local
curvature and torsion. A homogeneous, finite magnetic field,
concentric with the wedge dispiration line, is included. As a
working hypothesis, we assume the particle is bound in the
vicinity of the wedge dispiration by a planar potential given by a
long range attractive (harmonic) and short range (inverse square)
repulsive terms.

The paper is organized as follows: In Section II, we introduce the
topological defect. In Section III, we consider a simple finite
defect distribution describing a non-singular dispiration. In
Section IV,  the corresponding time-independent Schr\"{o}dinger
equation is taken into account. In Section V,  exact expressions
for the energy eigenfunctions of the particle moving both inside
and outside the defect core of the medium are obtained. Boundary
conditions on the surface of the defect region are implemented in
order to get the wavefunction defined over the whole space. The
energy spectrum is determined and we consider the limit in which
the defect region is shrunk to a null size. Conclusions are drawn
in Section VI.

\section{The Magnetic Wedge Dispiration}

In the framework of traditional elasticity theory, a Cartesian reference
frame $x^{i}$ is attached to the undistorted medium of an elastic solid with
Euclidean metric $\delta _{ab}$. The deformation of the medium is described
locally in terms of a continuous displacement function $u(x)$. In this way,
after a deformation has occurred, the point $x^{i}$ will have coordinates $%
x^{i}\rightarrow y^{i}\left( x\right) =x^{i}+u^{i}\left( x\right) $. The
initial metric $\delta _{ab}$ is transformed into \cite{Landau, 15}%
\begin{equation}
g_{ij}:=\frac{\partial x^{a}}{\partial y^{i}}\frac{\partial x^{b}}{\partial
y^{j}}\delta _{ab}\text{, }i\text{, }j=1\text{, }2\text{, }3\text{.}
\end{equation}%
We consider a topological defect in three dimensions whose geometry is
characterized by the spatial line element \cite{11}%
\begin{equation}
dl^{2}=g_{ij}dx^{i}dx^{j}=d\rho ^{2}+\kappa ^{2}\rho ^{2}d\varphi
^{2}+\left( dz+\tau d\varphi \right) ^{2}
\end{equation}%
where $\left( \rho \text{, }\varphi \text{, }z\right) $ are cylindrical
coordinates, with $\rho \geq 0$, $0\leq \varphi \leq 2\pi $, $-\infty \leq
z\leq \infty $, $\kappa $, $\tau \in
\mathbb{R}
$. The parameter $\kappa $ is related to the Frank vector $\vec{\Omega}$
\cite{KAT} of the disclination (describing curvature, i.e. the angular
deficit in the manifold) while $\tau $ is related to the Burgers' vector $%
\vec{b}$ \cite{Landau, KAT} of the dislocation (describing torsion). The
three dimensional geometry of the medium is therefore characterized by
nonvanishing torsion and curvature. When $\kappa \neq 0$ and $\tau =0$ we
have a wedge disclination; when $\kappa =0$ and $\tau \neq 0$ we have a
screw dislocation. The metric tensor $g_{ij}$ and its inverse $g^{ij}=\left(
g_{ij}\right) ^{-1}$ are given by \cite{14}%
\begin{equation}
g_{ij}=\left[
\begin{array}{ccc}
1 & 0 & 0 \\
0 & \kappa ^{2}\rho ^{2}+\tau ^{2} & \tau  \\
0 & \tau  & 1%
\end{array}%
\right] \text{, }g^{ij}=\left[
\begin{array}{ccc}
1 & 0 & 0 \\
0 & \frac{1}{\kappa ^{2}\rho ^{2}} & -\frac{\tau }{\kappa ^{2}\rho ^{2}} \\
0 & -\frac{\tau }{\kappa ^{2}\rho ^{2}} & 1+\frac{\tau ^{2}}{\kappa ^{2}\rho
^{2}}%
\end{array}%
\right] \text{.}
\end{equation}%
The line element $(2)$ describes an infinitely long linear wedge dispiration
oriented along the $z$-axis. We introduce the dual ($1$-form) basis vectors $%
\vartheta ^{i}=e_{\text{ }j}^{i}dx^{j}$ which describe the background $(2)$,
with coframe components \cite{15}%
\begin{equation}
\vartheta ^{1}\equiv \vartheta ^{\rho }=d\rho \text{, }\vartheta ^{2}\equiv
\vartheta ^{\varphi }=\kappa \rho d\varphi \text{ and }\vartheta ^{3}\equiv
\vartheta ^{z}=dz+\tau d\varphi \text{.}
\end{equation}%
The metric $g^{ij}$ and triad components $e_{\text{ }a}^{i}$ are related via
$e_{\text{ }a}^{i}e_{\text{ }b}^{j}\delta ^{ab}=g^{ij}$, where the triads
satisfy $e_{\text{ }j}^{a}e_{\text{ }b}^{j}=\delta _{b}^{a}$. In a
Riemannian geometry without torsion, the Ricci $R_{ij}$ and Riemann
curvature tensor $R_{\text{ }ijk}^{l}$ are given by \cite{MTW}%
\begin{equation}
R_{ij}=g^{ab}R_{aibj}=\partial _{k}\Gamma _{ij}^{k}-\partial _{j}\Gamma
_{ik}^{k}+\Gamma _{ij}^{k}\Gamma _{kn}^{n}-\Gamma _{ik}^{m}\Gamma _{jm}^{k}
\end{equation}%
and%
\begin{equation}
R_{\text{ }ijk}^{l}=\partial _{i}\Gamma _{jk}^{l}-\partial _{j}\Gamma
_{ik}^{l}+\Gamma _{im}^{l}\Gamma _{jk}^{m}-\Gamma _{jm}^{l}\Gamma _{ik}^{m}%
\text{,}
\end{equation}%
respectively. The Christoffel symbols $\Gamma _{kij}$ appearing in $(5)$ or $%
(6)$ are defined by \cite{MTW}%
\begin{equation}
\Gamma _{kij}:=\frac{1}{2}\left( \partial _{k}g_{ij}+\partial
_{i}g_{jk}-\partial _{j}g_{ki}\right)
\end{equation}%
where $\partial _{k}:=\partial /\partial x^{k}$. The nonvanishing components
of the Christoffel symbols in the space with metric $(3)$ are \cite{15}%
\begin{equation}
\Gamma _{\varphi \rho }^{z}=\Gamma _{\rho \varphi }^{z}=-\frac{\tau }{\rho }%
\text{, \ }\Gamma _{\varphi \varphi }^{\rho }=-\kappa ^{2}\rho \text{, \ }%
\Gamma _{\rho \varphi }^{\varphi }=\Gamma _{\varphi \rho }^{\varphi }=\frac{1%
}{\rho }\text{.}
\end{equation}%
By contrast, a non-Riemannian geometry is one that is characterized by
nonvanishing curvature and torsion. The torsion tensor $T_{iab}$ is defined
by%
\begin{equation}
T_{\text{ \ }ij}^{k}:=\partial _{i}e_{\text{ }j}^{k}-\partial _{j}e_{\text{ }%
i}^{k}+\Gamma _{i\nu }^{k}e_{\text{ }j}^{\nu }-\Gamma _{j\nu }^{k}e_{\text{ }%
i}^{\nu }\text{.}
\end{equation}%
The Ricci scalar is given by \cite{16, 17}%
\begin{equation}
R_{\text{ \ \ \ }12}^{12}=R_{\text{ \ }1}^{1}=R_{\text{ \ }2}^{2}=2\pi
\left( \frac{1-\kappa }{\kappa }\right) \delta ^{2}\left( \rho \right) \text{%
,}
\end{equation}%
where $\kappa =1+\phi /2\pi $ and $\delta ^{2}\left( \rho \right) $ is a
two-dimensional Dirac $\delta $-function. This $\delta $-function revels the
conic singularity in the curvature given in $(10)$. The dispiration
characterized by $(2)$ can be thought as arising from a "cut and paste"
process, known as the Volterra process \cite{14}. From the perspective of
the Volterra process, the disclination is generated by removing $(\kappa <1)$
or inserting $\left( \kappa >1\right) $ a wedge of material with deficit
angle $\phi =2\pi \left( \kappa -1\right) $. For $0<\kappa <1$ the
disclination carries positive curvature while for $1<\kappa <\infty $ it
carries negative curvature. The only non-vanishing component of torsion $2$%
-form $T^{k}=T_{\text{ \ }ij}^{k}dx^{i}\wedge dx^{j}$ is given by \cite{16,
17}%
\begin{equation}
T^{z}=2\pi \tau \delta ^{2}\left( \rho \right) d\rho \wedge d\varphi \text{.}
\end{equation}%
It is clear that the space described by metric $(3)$ features two conical
singularities at the origin as seen in $(10)$ and $(11)$. The Burger vector
can be viewed as a flux of torsion and the Frank vector as a flux of
curvature. The Burger vector is calculated by integrating around a closed
path $C$ encircling the dislocation \cite{Landau, 15}%
\begin{equation}
b^{3}=\oint\nolimits_{C}\vartheta ^{3}=\int_{S}d\rho \wedge d\varphi T_{%
\text{ \ }\rho \varphi }^{3}=2\pi \tau \text{,}
\end{equation}%
implying that $2\pi \tau $ is the flux intensity of the torsion source
passing through a closed loop $C$ in the $\widehat{e}_{z}$-direction. Thus,
the parameter $\tau $ is the modulus of the Burger vector. In a similar
manner, the Frank vector is determined by \cite{9, 15} $\vec{\Omega}%
:=\epsilon _{ijk}\Omega ^{jk}\widehat{e}_{i}$, with\
\begin{equation}
\Omega ^{12}=\int \int_{S}d\rho \wedge d\varphi R_{\text{ \ \ \ }\rho
\varphi }^{12}=2\pi \left( \frac{1-\kappa }{\kappa }\right)
\end{equation}%
where $\epsilon _{ijk}$ is the Levi-Civita symbol $(\epsilon _{123}=1)$, $%
\Omega ^{12}=-\Omega ^{21}$ and $S$ is a surface perpendicular to the defect
line, implying the flux of curvature is the surface density of the Frank
vector field.

\section{Finite Defect Distribution}

To avoid the singular nature of $(10)$ and $(11)$, we choose a simple
exactly solvable model of finite torsion and curvature. In the former case,
we choose a torsion field with a homogeneous flux distribution within the
dislocation region. In particular, we choose a torsion field specified by%
\begin{equation}
T^{z}=T^{z}\left( \rho \text{, }\varphi \right) =\tau ^{\prime }\Theta
\left( R_{c}-\rho \right) d\rho \wedge d\varphi \text{,}
\end{equation}%
where $R_{c}$ denotes the radius of the defect core, $\tau ^{\prime }=\frac{%
s_{T}b^{z}}{\pi R_{c}^{2}}$ and $s_{T}=\pm 1$ denotes the
handedness of the screw, where $(s_{T}=-1)$ indicates a left
handed screw in which a clockwise rotation of $\varphi =2\pi
\kappa ^{\prime }$ ($\kappa ^{\prime }$ is defined in Eq.$(17)$)
relative to the $\left( +\right) $ $x$-axis induces a shift in the
$\left( -\right) $ $z$-direction; similarly, $(s_{T}=+1)$
describes a right handed screw in which a counter-clockwise rotation of $%
\varphi =2\pi \kappa ^{\prime }$ relative to the $\left( +\right) $ $x$-axis
induces a shift in the $\left( +\right) $ $z$-direction. In $(14)$, $\Theta $
denotes the Heaviside theta function%
\begin{equation}
\Theta \left( R_{c}-\rho \right) =\left\{
\begin{array}{c}
1\text{ for }\rho <R_{c} \\
0\text{ for }\rho >R_{c}\text{.}%
\end{array}%
\right.
\end{equation}%
In the case of non-singular curvature, we consider a disclination
characterized by the deficit angle%
\begin{equation}
\overline{\phi }=\frac{1}{2}\phi R_{c}^{2}\Theta \left( R_{c}-\rho \right)
\text{.}
\end{equation}%
Under $\phi \rightarrow \overline{\phi }$ transformation, the angular
deficit $\kappa =1+\phi /2\pi $ transforms into
\begin{equation}
\kappa ^{\prime }=1+\frac{1}{2}\phi R_{c}^{2}\Theta \left( R_{c}-\rho
\right) =\left\{
\begin{array}{c}
\kappa _{\text{in}}=1+\frac{1}{2}\phi R_{c}^{2}\text{ for }\rho <R_{c} \\
\kappa _{\text{out}}=1\text{ for }\rho >R_{c}\text{.}%
\end{array}%
\right.
\end{equation}%
By the change of variables $\rho \rightarrow \overline{\rho }$, where%
\begin{equation}
\rho \rightarrow \overline{\rho }=\frac{\rho ^{\kappa ^{\prime }}}{\kappa
^{\prime }}=\left\{
\begin{array}{c}
\rho _{<}=\frac{\rho ^{\kappa _{\text{in}}}}{\kappa _{\text{in}}}\text{ for }%
\rho <R_{c} \\
\rho _{>}=\rho \text{ for }\rho >R_{c}\text{,}%
\end{array}%
\right.
\end{equation}%
the metric describing the non-singular wedge dispiration is given by \cite%
{16}%
\begin{equation}
dl^{2}=g_{ij}dx^{i}dx^{j}=d\overline{\rho }^{2}+\kappa ^{\prime 2}\overline{%
\rho }^{2}d\varphi ^{2}+\left( dz+\tau ^{\prime }d\varphi \right) ^{2}\text{.%
}
\end{equation}%
Observe that distributions $(14)$ and $(17)$ are chosen such that the total
torsion and curvature flux within the dispiration region are equivalent to
initial values $2\pi \tau $ and $2\pi \left( \frac{1-\kappa }{\kappa }%
\right) $ respectively. Furthermore, we assume the particle is moving in the
electromagnetic vector potential $\vec{A}\left( \overline{\rho }\right) $
\cite{Bordag}%
\begin{equation}
\vec{A}\left( \overline{\rho }\right) =\frac{B_{0}}{2\pi \kappa ^{\prime }%
\overline{\rho }}\left[ \frac{\overline{\rho }^{2}}{R_{c}^{2}}\Theta \left(
R_{c}-\rho \right) +\Theta \left( \rho -R_{c}\right) \right] \widehat{e}%
_{\varphi }
\end{equation}%
where $B_{0}$ is the total magnetic flux. Vector potential $\left( 20\right)
$ gives rise to a uniform magnetic field $\vec{B}$ within the dispirated
region.

\section{The Schr\"{o}dinger Equation}

We study the non-relativistic quantum dynamics of a charged spin-$\frac{1}{2}
$ particle propagating in a space with finite dispiration defect $(19)$ in
presence of a uniform magnetic field $\vec{B}=\vec{\nabla}\times \vec{A}$
and planar potential
\begin{equation}
V(\overline{\rho })=\frac{1}{2}M\omega ^{2}\overline{\rho }^{2}\Theta \left(
R_{c}-\rho \right) +\frac{f\hbar ^{2}}{2M\overline{\rho }^{2}}
\end{equation}%
where $f$ is a real, dimensionless constant characterizing the medium being
probed. Observe that the harmonic potential does not extend beyond the
medium $\left( R_{m}\right) $ being considered, where $R_{m}>>R_{c}$ and $%
R_{m}$ is the linear dimension of the medium. Furthermore, following \cite%
{20}, the spin-torsion interaction term $-\frac{1}{8}\vec{\sigma}\cdot \vec{T%
}$ is considered, where $\vec{\sigma}=\frac{g_{e}\mu }{\hbar }\widehat{e}%
_{S} $, $\widehat{e}_{S}$ denotes the spin direction, $\mu =\frac{q\hbar }{%
2Mc}$ is the Bohr magneton and $g_{e}$ is the gyromagnetic ratio of the
electron. Finally, the time-independent Schr\"{o}dinger equation becomes,%
\begin{equation}
\left[
\begin{array}{c}
\frac{1}{2M}\Delta +\vec{\sigma}\cdot \left( \vec{B}-\frac{1}{8}\vec{T}%
\right) +V\left( \overline{\rho }\right)%
\end{array}%
\right] \psi \left( \overline{\rho }\text{, }\varphi \text{, }z\right)
=E\psi \left( \overline{\rho }\text{, }\varphi \text{, }z\right) \text{.}
\end{equation}%
The wavefunction satisfying $(22)$ has periodicity $\psi \left( \overline{%
\rho }\text{, }\varphi \text{, }z\right) =\psi \left( \overline{\rho }\text{%
, }\varphi +2\pi \kappa ^{\prime }\text{, }z\right) $ rather than the usual
situation $\psi \left( \rho \text{, }\varphi \text{, }z\right) =\psi \left(
\rho \text{, }\varphi +2\pi \text{, }z\right) $ in flat space. The
Laplace-Beltrami operator $\Delta $ is defined by%
\begin{equation}
\Delta :=\frac{1}{\sqrt{g}}\left( \frac{\hbar }{i_{%
\mathbb{C}
}}\partial _{i}-\frac{q}{c}A_{i}\right) \left[ g^{ij}\sqrt{g}\left( \frac{%
\hbar }{i_{%
\mathbb{C}
}}\partial _{j}-\frac{q}{c}A_{j}\right) \right] =-\hbar ^{2}\vec{\nabla}^{2}+%
\frac{q^{2}}{c^{2}}\vec{A}^{2}-\frac{q\hbar }{ci_{%
\mathbb{C}
}}\left( \vec{\nabla}\cdot \vec{A}+\vec{A}\cdot \vec{\nabla}\right)
\end{equation}%
where $i_{%
\mathbb{C}
}$ is the imaginary unit, $c$ is the speed of light, $q$ is the charge of
the electron and $g=det\left\vert g_{ij}\right\vert $ is the determinant of
the metric. The Laplacian in the space $(2)$ is given by%
\begin{equation}
\vec{\nabla}^{2}=\left[ \frac{1}{\overline{\rho }}\frac{\partial }{\partial
\overline{\rho }}\left( \overline{\rho }\frac{\partial }{\partial \overline{%
\rho }}\right) +\frac{1}{\kappa ^{\prime 2}\overline{\rho }^{2}}\left( \frac{%
\partial }{\partial \varphi }-T^{z}\frac{\partial }{\partial z}\right) ^{2}+%
\frac{\partial ^{2}}{\partial z^{2}}\right] \text{.}
\end{equation}%
The divergence appearing in $(23)$ is computed according to%
\begin{equation}
\vec{\nabla}\cdot \vec{A}=\frac{1}{\overline{\rho }}\frac{\partial }{%
\partial \overline{\rho }}\left( \overline{\rho }A_{\overline{\rho }}\right)
+\frac{1}{\kappa ^{\prime }\overline{\rho }}\left( \frac{\partial }{\partial
\varphi }-T^{z}\frac{\partial }{\partial z}\right) A_{\varphi }+\frac{%
\partial A_{z}}{\partial z}
\end{equation}%
and is vanishing since $\vec{\nabla}\cdot \vec{A}=\frac{1}{\kappa ^{\prime }%
\overline{\rho }}\left( \frac{\partial }{\partial \varphi }-T^{z}\frac{%
\partial }{\partial z}\right) \frac{B_{0}}{2\pi \kappa ^{\prime }\overline{%
\rho }}=0$. With the aid of $(24)$ and $(25)$, operator $(23)$ can be
written as%
\begin{eqnarray}
\Delta &=&-\hbar ^{2}\vec{\nabla}^{2}+\frac{q^{2}}{c^{2}}\left( \frac{B_{0}}{%
2\pi \kappa ^{\prime }\overline{\rho }}\right) ^{2}\left[ \frac{\overline{%
\rho }^{2}}{R^{2}}\Theta \left( R_{c}-\rho \right) +\Theta \left( \rho
-R_{c}\right) \right] ^{2}+ \\
&&-\frac{qB_{0}}{2\pi c\kappa ^{\prime }\overline{\rho }}\frac{\hbar }{i_{%
\mathbb{C}
}}\left( \frac{\overline{\rho }^{2}}{R^{2}}\Theta \left( R_{c}-\rho \right)
+\Theta \left( \rho -R_{c}\right) \right) \left( \frac{\partial }{\partial
\varphi }-\frac{s_{T}b^{z}}{\pi R^{2}}\Theta \left( R_{c}-\rho \right) \frac{%
\partial }{\partial z}\right) \text{.}  \notag
\end{eqnarray}%
In the space $(2)$, the curl is given by%
\begin{eqnarray}
\vec{\nabla}\times \vec{A} &=&\left[ \frac{1}{\kappa ^{\prime }\overline{%
\rho }}\left( \frac{\partial }{\partial \varphi }-T^{z}\frac{\partial }{%
\partial z}\right) A_{z}-\frac{\partial }{\partial z}A_{\varphi }\right]
\widehat{e}_{\overline{\rho }}-\left[ \frac{\partial }{\partial \overline{%
\rho }}A_{z}-\frac{\partial }{\partial z}A_{\overline{\rho }}\right]
\widehat{e}_{\varphi }  \notag \\
&&+\frac{1}{\kappa \rho }\left[ \frac{\partial }{\partial \overline{\rho }}%
\left( \kappa ^{\prime }\overline{\rho }A_{\varphi }\right) -\left( \frac{%
\partial }{\partial \varphi }-T^{z}\frac{\partial }{\partial z}\right) A_{%
\overline{\rho }}\right] \widehat{e}_{z}\text{.}
\end{eqnarray}%
For the vector potential given in $(20)$, the magnetic field is computed
with the aid of $(27)$, the result being%
\begin{equation}
\vec{B}=\frac{B_{0}}{\kappa ^{\prime }\pi R_{c}^{2}}\Theta \left( R_{c}-\rho
\right) \widehat{e}_{z}
\end{equation}%
where $B_{0}$ is the total magnetic flux over the dispirated region with
radius $R_{c}$.

\section{Energy Eigenfunction and Eigenvalues}

The explicit time-independent Schr\"{o}dinger equation for the system being
considered reads%
\begin{equation}
\left\{ -\frac{\hbar ^{2}}{2M}\vec{\nabla}^{2}+\frac{1}{2M}\left( \frac{q^{2}%
}{c^{2}}\vec{A}^{2}-\frac{q\hbar }{ci_{%
\mathbb{C}
}}\vec{A}\cdot \vec{\nabla}\right) +\frac{1}{2}\mu g_{e}\left(
s_{B}\left\vert \vec{B}\right\vert -\frac{1}{8}s_{T}\left\vert \vec{T}%
\right\vert \right) +V(\overline{\rho })\right\} \psi =E\psi \text{,}
\end{equation}%
In $(29)$ we have made the replacement%
\begin{equation}
\vec{\sigma}\cdot \left( \vec{B}-\frac{1}{8}\vec{T}\right) =\frac{1}{2}\mu
g_{e}\left( s_{B}\left\vert \vec{B}\right\vert -s_{T}\frac{1}{8}\left\vert
\vec{T}\right\vert \right) =\frac{1}{2}\frac{q\hbar g_{e}}{2Mc}\left( s_{B}%
\frac{B_{0}}{\kappa ^{\prime }\pi R_{c}^{2}}-s_{T}\frac{1}{8}\frac{b^{z}}{%
\pi R_{c}^{2}}\right) \Theta \left( R_{c}-\overline{\rho }\right)
\end{equation}%
since both $\vec{B}$ and $\vec{T}$ are oriented along $\widehat{e}_{z}$,
where $s_{B}=\pm 1$ describes the orientation of the electron magnetic
moment relative to $\vec{B}$, either $(+)$ parallel or $(-)$ antiparallel.
In the region $\rho <R_{c}$, $\Theta \left( R_{c}-\rho \right) =1$ and $%
\Theta \left( \rho -R_{c}\right) =0$ resulting in the Schr\"{o}dinger
equation%
\begin{equation}
\left\{
\begin{array}{c}
-\frac{\hbar ^{2}}{2M}\left[ \frac{1}{\rho _{<}}\frac{\partial }{\partial
\rho _{<}}\left( \rho _{<}\frac{\partial }{\partial \rho _{<}}\right) +\frac{%
\partial ^{2}}{\partial z^{2}}+\frac{1}{\kappa _{\text{in}}^{2}\rho _{<}^{2}}%
\left( \frac{\partial }{\partial \varphi }-\tau ^{\prime }\frac{\partial }{%
\partial z}\right) ^{2}\right] \\
+\frac{1}{2M}\left[ \left( \frac{qB_{0}}{2\pi c\kappa _{\text{in}}R_{c}^{2}}%
\right) ^{2}\rho _{<}^{2}+\frac{i_{%
\mathbb{C}
}\hbar qB_{0}}{2\pi c\kappa _{\text{in}}^{2}R_{c}^{2}}\left( \frac{\partial
}{\partial \varphi }-\tau ^{\prime }\frac{\partial }{\partial z}\right) %
\right] \\
+\frac{1}{2}g_{e}\left( s_{B}\frac{B_{0}}{\kappa _{\text{in}}\pi R_{c}^{2}}%
-s_{T}\frac{1}{8}\frac{b^{z}}{\pi R_{c}^{2}}\right) \frac{q\hbar }{2Mc}+%
\frac{f\hbar ^{2}}{2M\overline{\rho }^{2}}+\frac{M\omega ^{2}\rho _{<}^{2}}{2%
}%
\end{array}%
\right\} \psi ^{\text{in}}=E\psi ^{\text{in}}\text{.}
\end{equation}%
In the region $\rho >R_{c}$, $\Theta \left( R_{c}-\rho \right) =0$ and $%
\Theta \left( \rho -R_{c}\right) =1$ yielding%
\begin{equation}
\left\{
\begin{array}{c}
-\frac{\hbar ^{2}}{2M}\left[ \frac{1}{\rho }\frac{\partial }{\partial \rho }%
\left( \rho \frac{\partial }{\partial \rho }\right) +\frac{\partial ^{2}}{%
\partial z^{2}}+\frac{1}{\rho ^{2}}\frac{\partial ^{2}}{\partial \varphi ^{2}%
}\right] + \\
+\frac{1}{2M}\left[ \left( \frac{qB_{0}}{2\pi c}\right) ^{2}\frac{1}{\rho
^{2}}+\frac{i_{%
\mathbb{C}
}\hbar qB_{0}}{2\pi c}\frac{1}{\rho ^{2}}\frac{\partial }{\partial \varphi }%
\right] +\frac{1}{2}M\omega ^{2}\rho ^{2}+\frac{f\hbar ^{2}}{2M\rho ^{2}}%
\end{array}%
\right\} \psi ^{\text{out}}=E\psi ^{\text{out}}\text{.}
\end{equation}%
In order to determine the eigenfunctions, the solutions to $(41)$ and $(42)$
must be matched at the boundary according to%
\begin{equation}
\psi ^{\text{in}}\left( \rho \right) |_{\rho =R_{c}}=\psi ^{\text{out}%
}\left( \rho \right) |_{\rho =R_{c}}\text{ and\ }\frac{\partial }{\partial
\rho }\psi ^{\text{in}}\left( \rho \right) |_{\rho =R_{c}}=\frac{\partial }{%
\partial \rho }\psi ^{\text{out}}\left( \rho \right) |_{\rho =R_{c}}\text{.}
\end{equation}%
Since we consider a scenario in which the particle is bound in the medium,
we require the wavefunctions satisfy the normalization condition%
\begin{equation}
\int_{0}^{R_{c}}\int\nolimits_{0}^{2\pi }\int\nolimits_{-\infty }^{\infty
}\left\vert \psi _{\text{in}}\right\vert ^{2}\rho d\rho d\varphi
dz+\int_{R_{c}}^{R_{m}}\int\nolimits_{0}^{2\pi }\int\nolimits_{-\infty
}^{\infty }\left\vert \psi _{\text{out}}\right\vert ^{2}\rho d\rho d\varphi
dz=1\text{.}
\end{equation}

\subsection{The Internal Core Solution}

Equation $(31)$ can be rewritten as%
\begin{equation}
\left\{
\begin{array}{c}
-\left[ \frac{1}{\rho _{<}}\frac{\partial }{\partial \rho _{<}}\left( \rho
_{<}\frac{\partial }{\partial \rho _{<}}\right) +\frac{\partial ^{2}}{%
\partial z^{2}}+\frac{1}{\kappa _{\text{in}}^{2}\rho _{<}^{2}}\left( \frac{%
\partial ^{2}}{\partial \varphi ^{2}}-\tau ^{\prime }\frac{\partial }{%
\partial z}\right) ^{2}\right] +\frac{M^{2}\omega ^{2}}{\hbar ^{2}}\rho
_{<}^{2}+\frac{f}{\rho _{<}^{2}} \\
+\frac{Q^{2}}{\kappa _{\text{in}}^{2}R_{c}^{4}}\rho _{<}^{2}+\frac{i_{%
\mathbb{C}
}}{\kappa _{\text{in}}^{2}R_{c}^{2}}Q\left( \frac{\partial }{\partial
\varphi }-\tau ^{\prime }\frac{\partial }{\partial z}\right) +g_{e}\Xi
-\varepsilon _{0}%
\end{array}%
\right\} \psi ^{\text{in}}=0\text{,}
\end{equation}%
where $Q=\frac{qB_{0}}{2\pi \hbar c}$, $\Xi =\frac{q}{\hbar c}\left( s_{B}%
\frac{B_{0}}{\kappa _{\text{in}}\pi R_{c}^{2}}-s_{T}\frac{1}{8}\frac{b^{z}}{%
\pi R_{c}^{2}}\right) $, $\varepsilon _{0}=\frac{2ME}{\hbar ^{2}}$. Since
the space $(2)$ possesses $\widehat{e}_{z}$-translational symmetry, we
assume a solution of the form $\psi ^{\text{in}}\left( \rho \text{, }\varphi
\text{, }z\right) =%
\mathbb{Z}
(z)\chi ^{\text{in}}\left( \rho \text{, }\varphi \right) $, with $%
\mathbb{Z}
(z)=e^{i_{%
\mathbb{C}
}k_{z}z}$. Moreover, since the cylindrical configuration being considered
possesses azimuthal symmetry, we assume $\chi ^{\text{in}}\left( \rho \text{%
, }\varphi \right) =\Phi \left( \varphi \right) \mathcal{R}^{\text{in}%
}\left( \rho \right) $, with $\Phi \left( \varphi \right) =e^{i_{%
\mathbb{C}
}l\varphi }$. Thus, $(35)$ reduces to%
\begin{equation}
\left\{ \rho _{<}^{2}\frac{d^{2}}{d\rho _{<}^{2}}+\rho _{<}\frac{d}{d\rho
_{<}}+\Lambda _{\text{in}}^{2}\rho _{<}^{2}-\Upsilon ^{2}\rho _{<}^{4}-\left[
f+\left( \frac{l-\tau ^{\prime }k_{z}}{\kappa _{\text{in}}}\right) ^{2}%
\right] \right\} \mathcal{R}^{\text{in}}\left( \rho \right) =0\text{,}
\end{equation}%
where $\Upsilon ^{2}=\left( \frac{Q}{R_{c}^{2}\kappa _{\text{in}}}\right)
^{2}+\frac{M^{2}\omega ^{2}}{\hbar ^{2}}$, $\Lambda _{\text{in}}^{2}=\frac{%
\left( l-\tau ^{\prime }k_{z}\right) Q}{R_{c}^{2}\kappa _{\text{in}}^{2}}%
+\varepsilon _{0}-k_{z}^{2}-g_{e}\Xi $. The solution to $(36)$ that is
regular at $\rho =0$\ is given in terms of confluent hypergeometric
functions $F$,%
\begin{equation}
\mathcal{R}^{\text{in}}\left( \rho \right) =A_{l}\rho _{<}^{\lambda }\exp
\left( -\frac{\Upsilon \rho _{<}^{2}}{2}\right) F\left( \frac{2-g_{e}}{4}+%
\frac{\lambda }{2}-\frac{\Lambda _{\text{in}}^{2}}{4\Upsilon }\text{, }%
1+\lambda \text{; }\Upsilon \rho _{<}^{2}\right)
\end{equation}%
where $A_{l}$ are appropriate expansion coefficients and $\lambda =\sqrt{%
\left( \frac{l-\tau ^{\prime }k_{z}}{\kappa _{\text{in}}}\right) ^{2}+f}$.

\subsection{The External Solution}

In the region $R_{c}<\rho $ we have%
\begin{equation}
\left\{ \frac{1}{\rho }\frac{\partial }{\partial \rho }\left( \rho \frac{%
\partial }{\partial \rho }\right) +\frac{\partial ^{2}}{\partial z^{2}}+%
\frac{1}{\rho ^{2}}\left( \frac{\partial }{\partial \varphi }-i_{%
\mathbb{C}
}Q\right) ^{2}-\frac{M^{2}\omega ^{2}}{\hbar ^{2}}\rho ^{2}-\frac{f}{\rho
^{2}}+\frac{2ME}{\hbar ^{2}}\right\} \psi ^{\text{out}}=0\text{.}
\end{equation}
Once again using a solution of the form $\psi ^{\text{out}}\left( \rho \text{%
, }\varphi \text{, }z\right) =%
\mathbb{Z}
(z)\Phi \left( \varphi \right) \mathcal{R}^{\text{out}}\left( \rho \right) $%
, with $%
\mathbb{Z}
(z)=e^{i_{%
\mathbb{C}
}k_{z}z}$ and $\Phi \left( \varphi \right) =e^{i_{%
\mathbb{C}
}l\varphi }$ we obtain%
\begin{equation}
\left\{ \rho ^{2}\frac{d^{2}}{d\rho ^{2}}+\rho \frac{d}{d\rho }+\varepsilon
_{\text{out}}^{2}\rho ^{2}-\varkappa ^{2}\rho ^{4}-\left[ \left( l+Q\right)
^{2}+f\right] \right\} \mathcal{R}^{\text{out}}\left( \rho \right) =0\text{,}
\end{equation}%
with $\varepsilon _{\text{out}}^{2}=\frac{2ME}{\hbar ^{2}}-k_{z}^{2}$, $%
\varkappa ^{2}=\frac{\omega ^{2}M^{2}}{\hbar ^{2}}$. The solution to $(39)$
that is regular at $\rho =0$\ is given by%
\begin{equation}
\mathcal{R}^{\text{out}}\left( \rho \right) =\exp \left( -\frac{\varkappa
\rho ^{2}}{2}\right) \left[
\begin{array}{c}
B_{l}\rho ^{\nu }F\left( \frac{1}{2}+\frac{\nu }{2}-\frac{\varepsilon _{%
\text{out}}^{2}}{4\varkappa }\text{, }1+\nu \text{; }\varkappa \rho
^{2}\right) \\
+C_{l}\rho ^{-\nu }F\left( \frac{1}{2}-\frac{\nu }{2}-\frac{\varepsilon _{%
\text{out}}^{2}}{4\varkappa }\text{, }1-\nu \text{; }\varkappa \rho
^{2}\right)%
\end{array}%
\right]
\end{equation}%
where $B_{l}$ and $C_{l}$ are expansion coefficients and $\nu =\sqrt{\left(
l+Q\right) ^{2}+f}$. Coefficients $A_{l}$, $B_{l}$ and $C_{l}$ are connected
via the boundary matching conditions $(33)$.

\subsection{Boundary Matching Conditions}

From the continuity of the wavefunction $\psi ^{\text{in}}\left( \rho
\right) |_{\rho =R_{c}}=\psi ^{\text{out}}\left( \rho \right) |_{\rho
=R_{c}} $ we obtain%
\begin{equation}
A_{l}=e^{\frac{\varkappa }{\Upsilon }}\left( B_{l}R_{c}^{\nu -\lambda
}M_{\nu \text{, }\lambda }+C_{l}R_{c}^{-\left( \nu +\lambda \right) }M_{-\nu
\text{, }\lambda }\right)
\end{equation}%
where%
\begin{equation}
M_{\nu \text{, }\lambda }:=\frac{F\left( \frac{1}{2}+\frac{\nu }{2}-\frac{%
\varepsilon _{\text{out}}^{2}}{4\varkappa }\text{, }1+\nu \text{; }\varkappa
R_{c}^{2}\right) }{F\left( \frac{2-g_{e}}{4}+\frac{\lambda }{2}-\frac{%
\Lambda _{\text{in}}^{2}}{4\Upsilon }\text{, }1+\lambda \text{; }\Upsilon
R_{c}^{2}\right) }\text{.}
\end{equation}%
To implement the second boundary condition in $(33)$ we make use of the
formula%
\begin{equation}
\frac{\partial F\left( a\text{, }b\text{; }z\right) }{\partial z}=\frac{a}{b}%
F\left( a+1\text{, }b+1\text{; }z\right) \text{.}
\end{equation}%
We compute the logarithmic derivative $\Pi ^{\text{in}}:=\rho \frac{1}{%
\mathcal{R}^{\text{in}}\left( \rho \right) }\frac{\partial }{\partial \rho }%
\mathcal{R}^{\text{in}}\left( \rho \right) |_{\rho =R_{c}}$, the result being%
\begin{equation}
\Pi ^{\text{in}}=A_{l}R_{c}\left( \lambda R_{c}^{-1}-\Upsilon
R_{c}+2\Upsilon R_{c}T_{\lambda }\right) \text{,}
\end{equation}%
where%
\begin{equation}
T_{\lambda }:=\frac{\frac{2-g_{e}}{4}+\frac{\lambda }{2}-\frac{\Lambda _{%
\text{in}}^{2}}{4\Upsilon }}{1+\lambda }\frac{F\left( 1+\frac{2-g_{e}}{4}+%
\frac{\lambda }{2}-\frac{\Lambda _{\text{in}}^{2}}{4\Upsilon }\text{, }%
1+1+\lambda \text{; }\Upsilon R_{c}^{2}\right) }{F\left( \frac{2-g_{e}}{4}+%
\frac{\lambda }{2}-\frac{\Lambda _{\text{in}}^{2}}{4\Upsilon }\text{, }%
1+\lambda \text{; }\Upsilon R_{c}^{2}\right) }\text{.}
\end{equation}%
In computing $(44)$ we made use of the relation%
\begin{equation}
\frac{d}{d\rho }F\left( \frac{2-g_{e}}{4}+\frac{\lambda }{2}-\frac{\Lambda _{%
\text{in}}^{2}}{4\Upsilon }\text{, }1+\lambda \text{; }\Upsilon \rho
^{2}\right) =2\Upsilon \rho \frac{\frac{2-g_{e}}{4}+\frac{\lambda }{2}-\frac{%
\Lambda _{\text{in}}^{2}}{4\Upsilon }}{1+\lambda }F\left( 1+\frac{2-g_{e}}{4}%
+\frac{\lambda }{2}\text{, }1+1+\lambda \text{; }\Upsilon \rho ^{2}\right)
\text{.}
\end{equation}%
From the logarithmic derivative of the external radial function, we obtain%
\begin{equation}
\Pi ^{\text{out}}=R_{c}B_{l}\left( \nu R_{c}^{-1}-\varkappa R_{c}+2\varkappa
R_{c}U_{\nu }\right) +R_{c}C_{l}\left( -\nu R_{c}^{-1}-\varkappa
R_{c}+2\varkappa R_{c}U_{-\nu }\right) \text{,}
\end{equation}%
where%
\begin{equation}
U_{\nu }:=\frac{\frac{1}{2}+\frac{\nu }{2}-\frac{\Lambda _{\text{out}}^{2}}{%
4a\varkappa }}{1+\nu }\frac{F\left( 1+\frac{1}{2}+\frac{\nu }{2}-\frac{%
\varepsilon _{\text{out}}^{2}}{4\varkappa }\text{, }1+1+\nu \text{; }%
\varkappa R_{c}^{2}\right) }{F\left( \frac{1}{2}+\frac{\nu }{2}-\frac{%
\varepsilon _{\text{out}}^{2}}{4\varkappa }\text{, }1+\nu \text{; }\varkappa
R_{c}^{2}\right) }\text{.}
\end{equation}%
From $(44)$ and $(47)$ we obtain%
\begin{equation}
A_{l}\left( \lambda -\Upsilon R_{c}^{2}+2\Upsilon R_{c}^{2}T_{\lambda
}\right) =B_{l}\left( \nu -\varkappa R_{c}^{2}+2\varkappa R_{c}^{2}U_{\nu
}\right) +C_{l}\left( -\nu -\varkappa R_{c}^{2}+2\varkappa R_{c}^{2}U_{-\nu
}\right) \text{.}
\end{equation}%
Eliminating the coefficient $A_{l}$ by substituting $(41)$ in $(49)$, the
matching condition reduces to%
\begin{equation}
\Upsilon R_{c}^{2}\left( \frac{\lambda }{\Upsilon R_{c}^{2}}-1+2T_{\lambda
}\right) =\frac{B_{l}\left( \nu R_{c}^{-1}-\varkappa R_{c}^{2}+2\varkappa
R_{c}^{2}U_{\nu }\right) +C_{l}\left( -\nu -\varkappa R_{c}^{2}+2\varkappa
R_{c}^{2}U_{-\nu }\right) }{B_{l}R_{c}^{\nu -\lambda }M_{\nu \text{, }%
\lambda }\exp \left[ \frac{R_{c}^{2}}{2}\left( \Upsilon -\varkappa \right) %
\right] +C_{l}R_{c}^{-\left( \nu +\lambda \right) }M_{-\nu \text{, }\lambda
}\exp \left[ \frac{R_{c}^{2}}{2}\left( \Upsilon -\varkappa \right) \right] }%
\text{.}
\end{equation}%
Equation $(50)$ can be put into the form%
\begin{equation}
B_{l}=C_{l}\aleph _{\nu \text{, }\lambda }^{(1)}
\end{equation}%
where%
\begin{equation}
\aleph _{\nu \text{, }\lambda }^{(1)}:=\frac{-\nu -\varkappa
R_{c}^{2}+2\varkappa R_{c}^{2}U_{-\nu }-\exp \left[ \frac{R_{c}^{2}}{2}%
\left( \Upsilon -\varkappa \right) \right] R_{c}^{-\left( \nu +\lambda
\right) }M_{-\nu \text{, }\lambda }\left( \lambda -\Upsilon
R_{c}^{2}+2\Upsilon R_{c}^{2}T_{\lambda }\right) }{\exp \left[ \frac{%
R_{c}^{2}}{2}\left( \Upsilon -\varkappa \right) \right] R_{c}^{\nu -\lambda
}M_{\nu \text{, }\lambda }\left( \lambda -\Upsilon R_{c}^{2}+2\Upsilon
R_{c}^{2}T_{\lambda }\right) -\left( \nu R_{c}^{-1}-\varkappa
R_{c}^{2}+2\varkappa R_{c}^{2}U_{\nu }\right) }\text{.}
\end{equation}%
Substituting $(51)$ into $(41)$ we obtain%
\begin{equation}
C_{l}=A_{l}\aleph _{\nu \text{, }\lambda }^{(2)}\text{,}
\end{equation}%
where%
\begin{equation}
\aleph _{\nu \text{, }\lambda }^{(2)}:=\frac{\lambda -\Upsilon
R_{c}^{2}+2\Upsilon R_{c}^{2}T_{\lambda }}{\aleph _{\nu \text{, }\lambda
}^{(1)}\left( \nu -\varkappa R_{c}^{2}+2\varkappa R_{c}^{2}U_{\nu }\right)
+\left( -\nu -\varkappa R_{c}^{2}+2\varkappa R_{c}^{2}U_{-\nu }\right) }%
\text{.}
\end{equation}%
With the aid of $(53)$, $(51)$ can be rewritten as%
\begin{equation}
B_{l}=A_{l}\aleph _{\nu \text{, }\lambda }^{(2)}\aleph _{\nu \text{, }%
\lambda }^{(1)}\text{.}
\end{equation}%
Conditions $(53)$ and $(55)$ determines the wavefunction up to a
normalization constant. In particular, the wavefunction becomes%
\begin{eqnarray}
&&\psi _{nl}\left( \rho \text{, }\varphi \text{, }z\right) \\
&=&A_{l}e^{i_{%
\mathbb{C}
}l\varphi }e^{i_{%
\mathbb{C}
}k_{z}z}\left\{
\begin{array}{c}
\rho ^{\lambda }\exp \left( -\frac{\Upsilon \rho _{<}^{2}}{2}\right) F\left(
\frac{2-g_{e}}{4}+\frac{\lambda }{2}-\frac{\Lambda _{\text{in}}^{2}}{%
4\Upsilon }\text{, }1+\lambda \text{; }\Upsilon \rho _{<}^{2}\right) \text{
for }\rho <R_{\text{core}}\text{ (in)} \\
\aleph _{\nu \text{, }\lambda }^{(2)}\exp \left( -\frac{\varkappa \rho ^{2}}{%
2}\right) \left[
\begin{array}{c}
\aleph _{\nu \text{, }\lambda }^{(1)}\rho ^{\nu }F\left( \frac{1}{2}+\frac{%
\nu }{2}-\frac{\varepsilon _{\text{out}}^{2}}{4\varkappa }\text{, }1+\nu
\text{; }\varkappa \rho ^{2}\right) + \\
+\rho ^{-\nu }F\left( \frac{1}{2}-\frac{\nu }{2}-\frac{\varepsilon _{\text{%
out}}^{2}}{4\varkappa }\text{, }1-\nu \text{; }\varkappa \rho ^{2}\right)%
\end{array}%
\right] \text{ for }\rho >R_{\text{core}}\text{ (out).}%
\end{array}%
\right.  \notag
\end{eqnarray}%
The asymptotic expansion of the confluent hypergeometric function is given by%
\begin{equation}
F\left( a\text{, }c\text{; }x\right) \simeq \frac{\Gamma \left( c\right) }{%
\Gamma \left( a\right) }\left\vert x\right\vert ^{a-c}e^{i_{%
\mathbb{C}
}\left( a-c\right) \varphi }e^{\left\vert x\right\vert ^{e^{i_{%
\mathbb{C}
}\varphi }}}+\frac{\Gamma \left( c\right) }{\Gamma \left( c-a\right) }%
\left\vert x\right\vert ^{-a}e^{i_{%
\mathbb{C}
}a\left( \pi -\varphi \right) }\text{.}
\end{equation}%
If $x=i_{%
\mathbb{C}
}z$ ($z=\left\vert x\right\vert $ - real, positive), then $\varphi =\pi /2$\
and both terms in $(57)$ are approximately equally large and should be taken
into account. Provided $x$ is real, positive $(\varphi =0)$, the first term
is considered only. When $x$ is real, negative $(\varphi =\pi )$ the second
term on the right hand side of $(57)$ is used \cite{Feshbach}. The radial
function $(37)$ is ensured to be vanishing at the origin $\rho =0$ by
expanding $\left( 37\right) $ according to $\left( 57\right) $ and choosing%
\begin{equation}
\frac{2-g_{e}}{4}+\frac{1}{2}\sqrt{\left( \frac{l-\tau ^{\prime }k_{z}}{%
\kappa _{\text{in}}}\right) ^{2}+f}-\frac{\Lambda _{\text{in}}^{2}}{%
4\Upsilon }=-n\text{ where }n\in \mathbb{Z}
\end{equation}%
which corresponds to the pole in the second term of $(57)$. From $(58)$ the
energy eigenvalues in the defect region can be written as%
\begin{eqnarray}
E_{nl}^{\text{in}} &=&2\hbar \left( \frac{\omega _{B}^{2}}{4}+\omega
^{2}\right) ^{1/2}\left[ n+\sqrt{\left( \frac{l-\tau ^{\prime }k_{z}}{%
2\kappa _{\text{in}}}\right) ^{2}+\frac{f}{4}}+\frac{2-g_{e}}{4}\right] -%
\frac{\hbar \omega _{B}}{2}\frac{l-\tau ^{\prime }k_{z}}{\kappa _{\text{in}}}
\notag \\
&&+\frac{q\hbar g_{e}}{2Mc}\left( s_{B}\left\vert \vec{B}\right\vert -s_{T}%
\frac{1}{8}\left\vert \vec{T}\right\vert \right) +\frac{\hbar ^{2}k_{z}^{2}}{%
2M}\text{,}
\end{eqnarray}%
where $\omega _{B}:=qB_{0}/\pi \kappa _{\text{in}}R_{c}^{2}Mc=q\left\vert
\vec{B}\right\vert /Mc$ is the Larmor frequency of the electron and we made
use of $\Upsilon =\sqrt{\left( \frac{M\omega _{B}}{2\hbar }\right) ^{2}+%
\frac{M^{2}\omega ^{2}}{\hbar ^{2}}}$. Similarly, demanding $(40)$ be
vanishing as $\rho \rightarrow \infty $ requires that we choose%
\begin{equation}
\frac{1}{2}+\frac{1}{2}\sqrt{\left( l+Q\right) ^{2}+f}-\frac{\varepsilon _{%
\text{out}}^{2}}{4\varkappa }=-n
\end{equation}%
in the expansion of the confluent hypergeometric function whose coefficient
is $B_{l}$. From $(60)$ the energy eigenvalues external to the defect region
can be written as%
\begin{equation}
E_{nl}^{\text{out}}=2\hbar \omega \left( n+\sqrt{\left( \frac{l+Q}{2}\right)
^{2}+\frac{f}{4}}+\frac{1}{2}\right) +\frac{\hbar ^{2}k_{z}^{2}}{2M}\text{.}
\end{equation}

\subsection{The Limit $R_{c}\rightarrow 0$}

In order to facilitate a comparison between our results and the
Aharonov-Bohm type scenario, we consider the limit in which the defect
region is shrunk to zero, that is $R_{c}\rightarrow 0$. In this limit, the
dispiration along with the internal magnetic field is shrunk to an
infinitesimal line along the $z$-axis. In this case the spin-magnetic and
spin-torsion interaction vanishes, $T^{z}$ is given by $(11)$, $R$ by $(10)$
and $V(\rho )=\frac{M\omega ^{2}\rho ^{2}}{2}+\frac{f\hbar ^{2}}{2M\rho ^{2}}
$. The corresponding Schr\"{o}dinger equation reads%
\begin{equation}
\left\{
\begin{array}{c}
-\left[ \frac{1}{\rho }\frac{\partial }{\partial \rho }+\frac{\partial ^{2}}{%
\partial \rho ^{2}}+\frac{\partial ^{2}}{\partial z^{2}}+\frac{1}{\kappa
^{2}\rho ^{2}}\left( \frac{\partial }{\partial \varphi }-\tau \frac{\partial
}{\partial z}\right) ^{2}\right] -\frac{1}{i_{%
\mathbb{C}
}}\frac{qB_{0}}{2\pi \hbar c\kappa ^{2}\rho ^{2}}\left( \frac{\partial }{%
\partial \varphi }-\tau \frac{\partial }{\partial z}\right) \\
+\left( \frac{qB_{0}}{2\pi \hbar c\kappa \rho }\right) ^{2}+\frac{%
M^{2}\omega ^{2}}{\hbar ^{2}}\rho ^{2}+\frac{f\hbar ^{2}}{\hbar ^{2}\rho ^{2}%
}%
\end{array}%
\right\} \psi =E\psi \text{.}
\end{equation}%
With a solution of form $\psi \left( \rho \text{, }\varphi \text{, }z\right)
=%
\mathbb{Z}
(z)\Phi \left( \varphi \right) \mathcal{R}\left( \rho \right) $ where $%
\mathbb{Z}
(z)=e^{i_{%
\mathbb{C}
}k_{z}z}$ and $\Phi \left( \varphi \right) =e^{i_{%
\mathbb{C}
}l\varphi }$, equation $(62)$ is transformed into%
\begin{equation}
\left\{ \rho ^{2}\frac{d^{2}}{d\rho ^{2}}+\rho \frac{d}{d\rho }-\left[ \frac{%
\left( l-\tau k_{z}\right) ^{2}}{\kappa ^{2}}+f\right] +\frac{1}{\kappa ^{2}}%
\left[ \frac{qB_{0}}{2\pi \hbar c}\left( l-\tau k_{z}\right) -\left( \frac{%
qB_{0}}{2\pi \hbar c}\right) ^{2}\right] +\varkappa \rho ^{4}+\Lambda
^{2}\rho ^{2}\right\} \mathcal{R}\left( \rho \right) =0\text{,}
\end{equation}%
with $\Lambda ^{2}=\frac{2ME}{\hbar ^{2}}-k\ $and $\varkappa ^{2}=\frac{%
M^{2}\omega ^{2}}{\hbar ^{2}}$. The solution to $(63)$ that is regular at $%
\rho =0$ is given by%
\begin{equation}
\mathcal{R}\left( \rho \right) =\rho ^{\mu }\exp \left( -\frac{\varkappa
\rho ^{2}}{2}\right) F\left( \frac{1}{2}+\frac{\mu }{2}-\frac{\Lambda ^{2}}{%
4\varkappa }\text{, }1+\mu \text{; }\varkappa \rho ^{2}\right) \text{,}
\end{equation}%
where $\mu =\sqrt{\left( \frac{l+Q-\tau k_{z}}{\kappa }\right) ^{2}+f}$. In
order that $(64)$ be vanishing as $\rho \rightarrow \infty $, we require%
\begin{equation}
\frac{1}{2}+\frac{1}{2}\sqrt{\left( \frac{l+Q-\tau k_{z}}{\kappa }\right)
^{2}+f}-\frac{\Lambda ^{2}}{4\varkappa }=-n
\end{equation}%
which leads to%
\begin{equation}
E_{nl}^{R_{c}\rightarrow 0}=2\hbar \omega \left( n+\sqrt{\left( \frac{%
l+Q-\tau k_{z}}{2\kappa }\right) ^{2}+\frac{f}{4}}+\frac{1}{2}\right) +\frac{%
\hbar ^{2}k_{z}^{2}}{2M}\text{.}
\end{equation}%
Finally, the eigenfunction associated to the Aharonov-Bohm type scenario
associated with the magnetic wedge dispiration reads%
\begin{equation}
\psi _{nl}^{R_{c}\rightarrow 0}\left( \rho \text{, }\varphi \text{, }%
z\right) =A_{l}e^{i_{%
\mathbb{C}
}k_{z}z}e^{i_{%
\mathbb{C}
}l\varphi }\rho ^{\mu }\exp \left( -\frac{\varkappa \rho ^{2}}{2}\right)
F\left( -n\text{, }1+\mu \text{; }\varkappa \rho ^{2}\right)
\end{equation}%
where $A_{l}$ is a normalization constant. We observe the modification of
the angular momentum $l\rightarrow l^{\prime }=\left( l+Q-\tau k_{z}\right)
/\kappa $, interpreted as an extension of the Aharonov-Bohm effect
accounting for the influence of the magnetic flux $Q$ (via the vector
potential $\vec{A}$) and the topological defect (via the curvature $\kappa $
and torsion $\tau $ parameters).

\section{Final Remarks}

In this paper, we obtained exact expressions for the eigenvalues and
eigenfunctions of a charged particle with magnetic moment bound in the
vicinity of a magnetic wedge dispiration by a short range repulsive and long
range attractive potentials. The screw dislocation modified the angular
momentum by introducing an additive correction in a similar manner as the
magnetic flux. The disclination introduced a multiplicative modification to
the angular momentum, the appearance of which is understood as a consequence
of the modified periodicity of the wavefunction in the space of the wedge
dispiration around the $z$-axis. Furthermore, due to the finite size of the
defect we were able to account for the effects of spin-torsion and
spin-magnetic field interactions. The handedness of the screw dislocation
simulates the north/south pole of the magnetic field, increasing(decreasing)
the binding energy of the particle in the core region for left(right) handed
screws. In the limit where the defect region is shrunk to zero radius, the
angular momentum is modified so as to depend not only on the magnetic flux,
but also on the screw dislocation and wedge disclination parameters. Being
that the defects characterized by parameters $\kappa $ and $\tau $ are
singular at the defect line and vanishing elsewhere in the limit $%
R_{c}\rightarrow 0$, the appearance of global phenomena represented by the
change in angular momentum is interpreted as a manifestation of the
topological features of the space rather than the local geometry induced by
the defect.

\end{document}